\documentclass[9pt,twocolumn,twoside]{optica_arxiv}

\setboolean{minireview}{false}
\setboolean{displaycopyright}{false}
\doi{}

\usepackage{amsmath, mathrsfs}
\usepackage{xspace}
\usepackage{xr}

\usepackage{booktabs}
\usepackage[capitalise]{cleveref}
\usepackage{todonotes}
\usepackage{dblfloatfix}

\DeclareMathOperator{\N}{\mathbb{N}}

\newcommand{\sizethree}[3]{\ensuremath{#1\mkern-2mu\times\mkern-2mu#2\mkern-2mu\times\mkern-2mu#3\xspace}}

\makeatletter
\DeclareRobustCommand\onedot{\futurelet\@let@token\@onedot}
\def\@onedot{\ifx\@let@token.\else.\null\fi\xspace}

\def\eg{\emph{e.g}\onedot} 
\def\ie{\emph{i.e}\onedot} 
\def\cf{\emph{cf.}\xspace}

 \def\vs{\emph{vs}\onedot\xspace}

\makeatother

\newcommand{\um}{\ensuremath{\mu m}\xspace}
\newcommand{\nm}{\ensuremath{nm}\xspace}

\newcommand{\NA}{\mathit{NA}}

\newcommand{\phasenet}{\textsc{PhaseNet}\xspace}
\newcommand{\Zola}{\textsc{Zola}\xspace}

\newcommand{\sted}{\textsc{Point~scanning}\xspace}
\newcommand{\wide}{\textsc{Widefield}\xspace}

\newcommand{\SM}[1]{Supp.~Notes~\ref*{#1}}
\newcommand{\ST}[1]{Supp.~Table~\ref*{#1}}
\newcommand{\SF}[1]{Supp.~Fig.~\ref*{#1}}

\title{Practical sensorless aberration estimation for 3D microscopy with deep learning}

\author[1,2]{Debayan Saha}
\author[1,2]{Uwe Schmidt}
\author[3]{Qinrong Zhang}
\author[4]{Aurelien Barbotin}
\author[4]{Qi Hu}
\author[3]{Na Ji}
\author[4,*]{Martin J. Booth}
\author[1,2,5,*]{Martin Weigert}
\author[1,2,*]{Eugene W. Myers}

\affil[1]{MPI-CBG, Dresden, Germany}
\affil[2]{CSBD, Dresden, Germany}
\affil[3]{University of California, Berkeley, USA}
\affil[4]{University of Oxford, Department of Engineering Science, Oxford, UK}
\affil[5]{Institute of Bioengineering, School of Life Sciences, EPFL, Lausanne, Switzerland}

\affil[*]{Corresponding author: myers@mpi-cbg.de, martin.weigert@epfl.ch, martin.booth@eng.ox.ac.uk}

\begin{abstract}
Estimation of optical aberrations from volumetric intensity images is a key step in sensorless adaptive optics for 3D microscopy.
Recent approaches based on deep learning promise accurate results at fast processing speeds. 
However, collecting ground truth microscopy data for training the network is typically very difficult or even impossible thereby limiting this approach in practice.
Here, we demonstrate that neural networks trained \emph{only} on \emph{simulated} data yield accurate predictions for \emph{real} experimental images.
We validate our approach on simulated \emph{and} experimental datasets acquired with two different microscopy modalities, and also compare the results to non-learned methods.
Additionally, we study the predictability of individual aberrations with respect to their data requirements and find that the symmetry of the wavefront plays a crucial role.
Finally, we make our implementation freely available as open source software in Python.
 \end{abstract}

\begin{document}

\maketitle

\newcommand{\figOverview}{{
\begin{figure}[t]
  \centering
    \includegraphics[width=1.\linewidth]{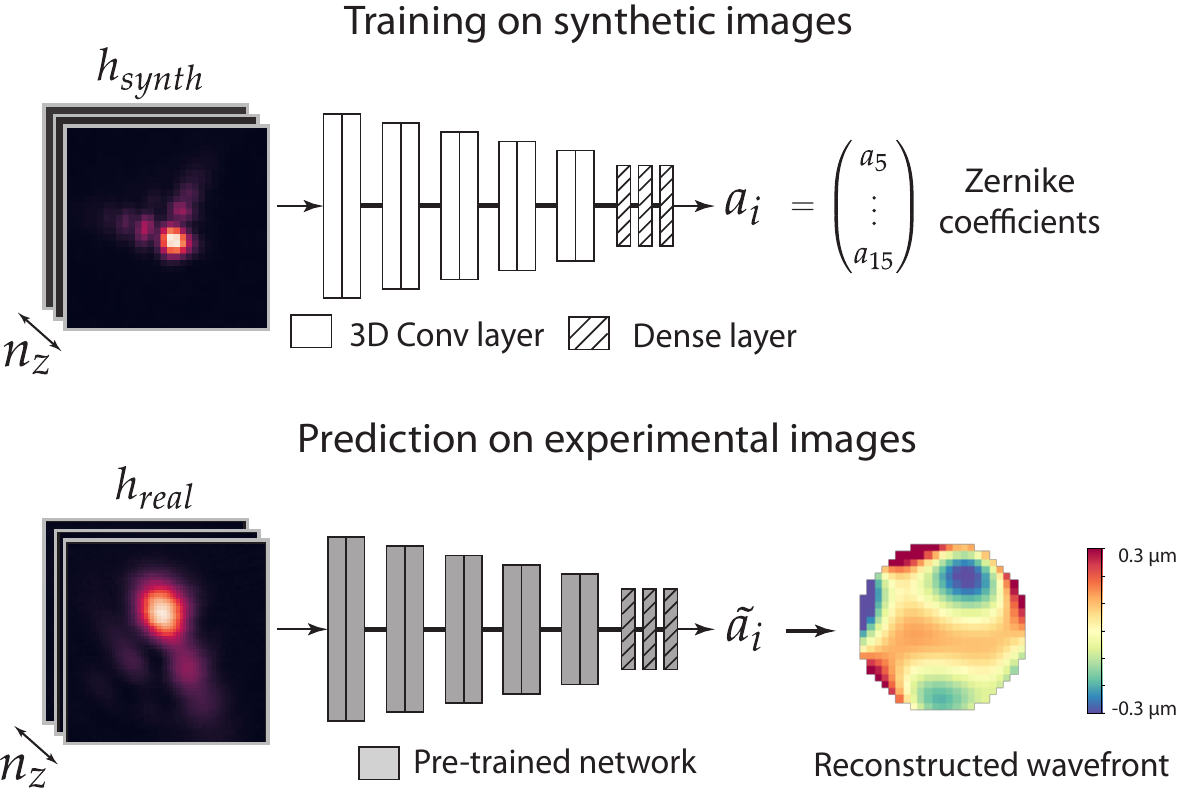}
    \caption{Overview of our approach: We train a CNN (\phasenet) with synthetic PSFs $h_{synth}$ ($n_{z}$ axial planes) generated from randomly sampled amplitudes of Zernike modes $a_i$.
 The trained network is then used to predict the amplitudes $\tilde{a}_i$ from experimental bead images $h_{real}$. The predicted amplitudes $\tilde{a}_i$ are then used to reconstruct the wavefront.}
\label{fig:overview}
\end{figure}
}}

\newcommand{\figResultsSingle}{{
\begin{figure*}[t]
  \centering
    \includegraphics[width=1\linewidth]{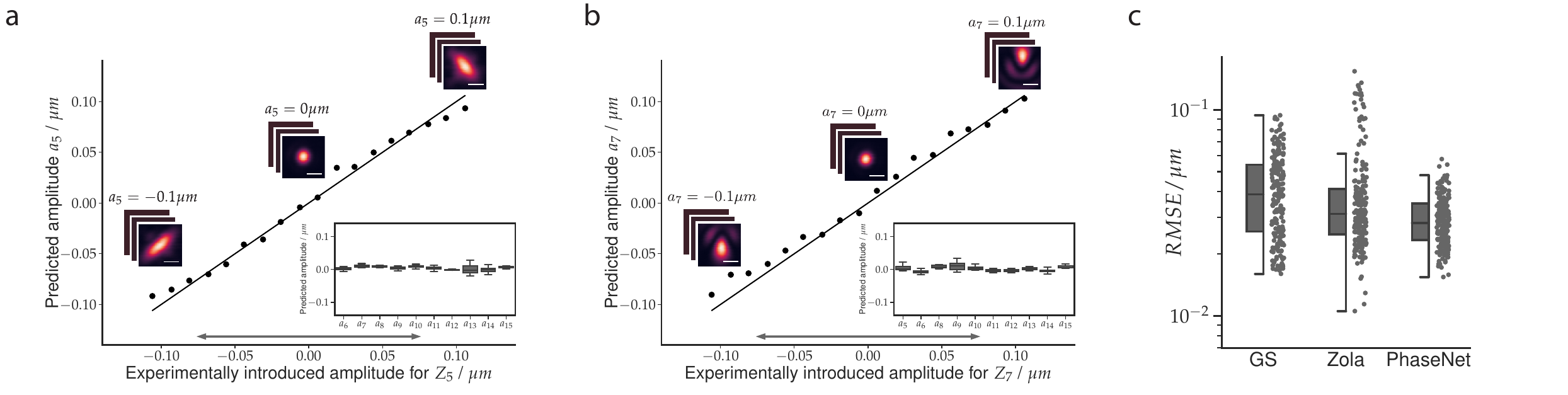}
    \caption{Measurement of single Zernike mode aberrations for \sted data. a)
      \phasenet predictions on images with experimentally introduced oblique astigmatism $Z_5$ (see \SF{fig:supp:stedsynth} for modes $Z_6$ - $Z_{15}$).
      Shown are ground truth \vs the predicted amplitude $a_5$ (black dots), perfect prediction (solid black line), and the upper/lower bounds of amplitudes used during training (gray arrow). 
      The inset shows the distribution of predicted non-introduced modes $(a_{6}, \dots, a_{15})$.
      b) Same results for experimentally introduced vertical coma $Z_7$. 
       Scalebar $500\nm$.
      c) RMSE for \phasenet and compared methods (GS and \Zola) on all images. Boxes show interquartile range (IQR), lines signify median, and whiskers extend to 1.5 IQR.
    }
\label{fig:results_single}
\end{figure*}
}}

\newcommand{\figResultsRandom}{{
\begin{figure*}[t]
  \centering
    \includegraphics[width=1\linewidth]{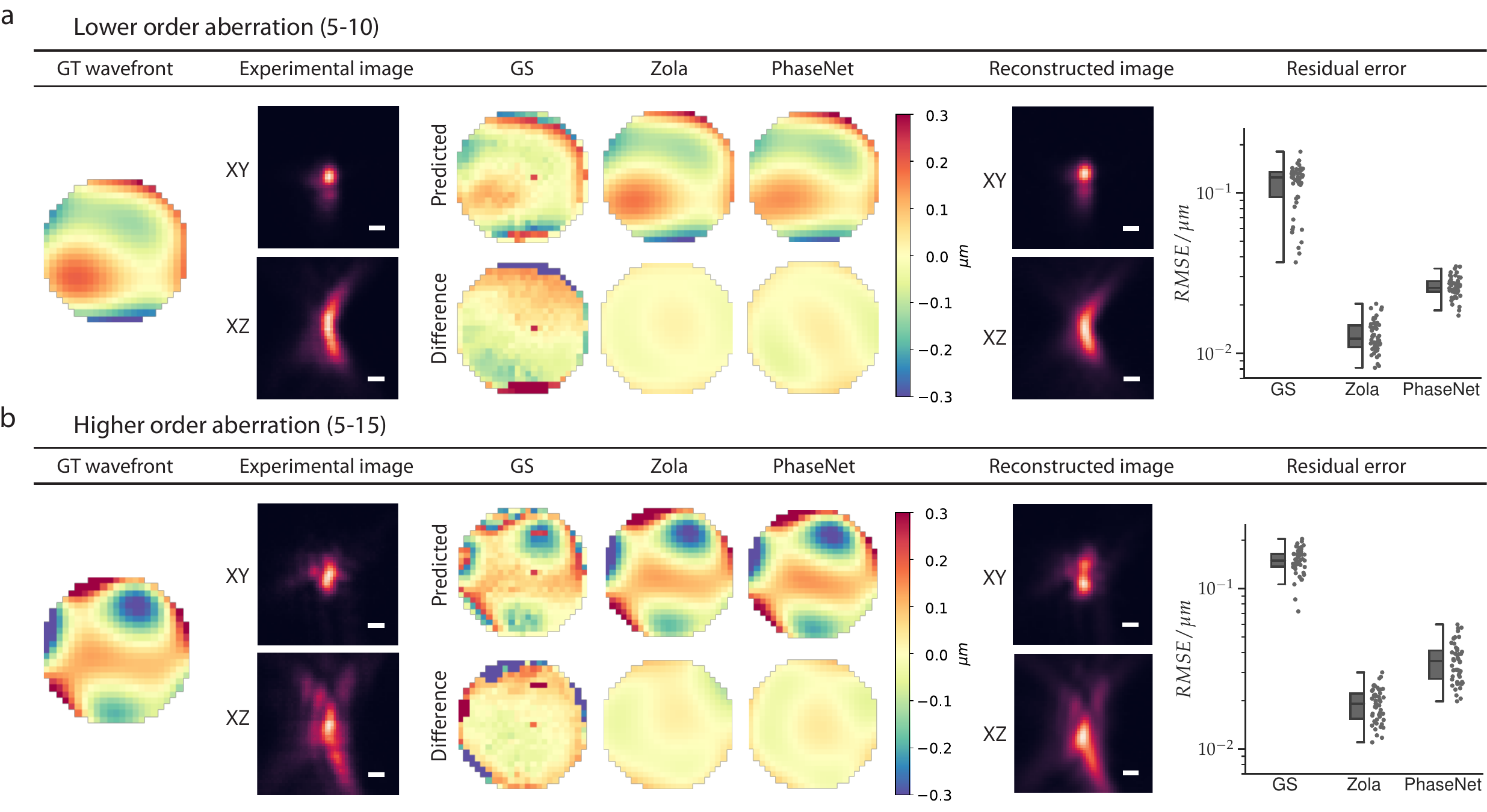}
    \caption{Results for \wide data with mixed-modes aberrations. a) Predictions for lower order modes ($Z_5 - Z_{10}$):
      We show the ground truth (GT) wavefront, lateral (XY) and axial (XZ) midplanes of the experimental 3D image, the reconstructed wavefront and their GT difference for all methods (Gerchberg-Saxton/GS~\cite{Kner2010}, Zola~\cite{Aristov2018}, \phasenet), and the reconstructed image from the \phasenet prediction.
      We further depict the RMSE for all $n=50$ experimental PSFs. Boxes show interquartile range (IQR), lines signify median, and whiskers extend to 1.5 IQR. 
      b) Same results but including higher order modes $Z_5 - Z_{15}$. Scalebar: 500 \nm.}
\label{fig:results_random}
\end{figure*}
}}

\newcommand{\figResultsPlanes}{
\begin{figure*}[t]
  \centering
  \includegraphics[width=1\linewidth]{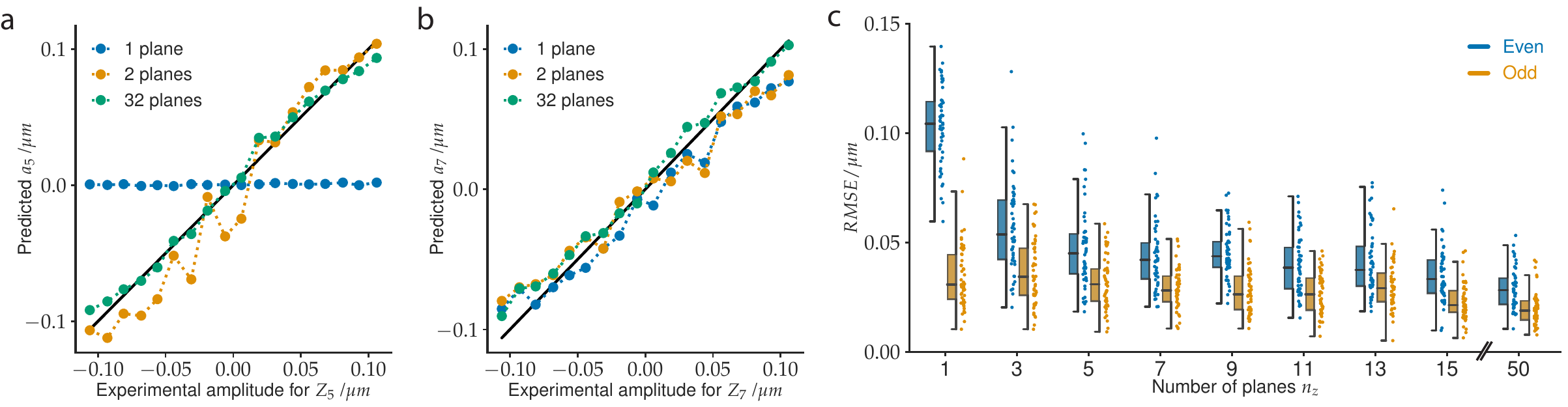}
  \caption{Results for varying number of input planes $n_z$: a) Ground truth \vs the predicted amplitude $a_5$ (oblique astigmatism) for single mode data \sted and using \phasenet models with $n_z=1, 2, 32$. b) The same for $a_7$ (vertical coma). c) Prediction error (RMSE) on \wide data (50 images) for \phasenet models trained with different $n_z$. We show the RMSE for \emph{odd} (orange) and \emph{even} (blue) Zernike modes separately. Boxes depict interquartile range (IQR), lines signify median, and whiskers extend to 1.5 IQR.}
\label{fig:results_planes}
\end{figure*}
}

\section{Introduction}

Image quality in volumetric microscopy of biological samples is often severely limited by optical aberrations due to refractive index inhomogeneities inside the specimen~\cite{Schwertner2004, Kubby2013}.
Adaptive optics (AO) is widely used to correct for these distortions via optical elements like deformable mirrors or spatial light modulators~\cite{Booth2014, Ji2017}.
Successful implementation of AO requires aberration measurement at multiple locations within the imaging volume~\cite{Liu2018}.
This can be achieved by creating point sources such as embedded fluorescent beads~\cite{Ji2012} or optically induced guide stars~\cite{Wang2015}, and then sensing the wavefront either \emph{directly} via dedicated hardware (\eg Shack-Hartman wavefront sensors~\cite{cha2010,Kubby2012}) or \emph{indirectly} from the intensity image of the point source (PSF) alone~\cite{Fienup1982, Kner2010}.
Due to its special hardware requirements, and its reliance on a point-scanning configuration, direct wavefront sensing can be cumbersome to implement and too slow for volumetric imaging of living samples~\cite{booth2007adaptive}.
In contrast, indirect wavefront sensing - or \emph{phase retrieval} - offers the possibility to infer the aberration at multiple locations, across the entire volume simultaneously, without additional optical hardware~\cite{debarre2009, xu2020}.
Establishing a fast and accurate phase retrieval method from intensity images of point sources is therefore an important step for making AO more accessible to live imaging of large biological samples.

Classical approaches to phase retrieval include alternating projection methods such as Gerchberg-Saxton (GS)~\cite{hanser2003,Kner2010} or parameterized PSF fitting methods such as \Zola~\cite{Aristov2018}. 
While projection methods are typically fast but can perform poorly especially for noisy images, PSF fitting methods can achieve excellent results yet are relatively slow.
Over the last years, deep learning based approaches using convolutional neural networks (CNNs) have proven to be powerful and computationally efficient for image-based classification and regression tasks for microscopy images~\cite{rivenson2017, Weigert2018}.
Recently, several studies demonstrated that deep learning based phase retrieval can produce accurate results at fast processing speeds~\cite{Zhang2018, Jin18, paine2018, moeckl2019}, however they fall short regarding their practical applicability.
Some of these approaches~\cite{paine2018, moeckl2019} used purely simulated synthetic data, where generalizability to real microscopy images is unclear.
Others~\cite{Zhang2018, Jin18} focused on specific microscopy modalities where it is feasible to collect large sets of experimental ground truth data for training and prediction, thus limiting this approach in practice.
Moreover, most studies lack comparison against strong classical phase retrieval methods that are used in practice.   
As a result, the practical applicability of these approaches in experimental microscopy settings remains unclear.

In this paper we demonstrate that CNNs trained on appropriately generated synthetic data can be successfully applied to real images acquired with different microscopy modalities thereby avoiding the cumbersome collection of experimental training data.
Specifically, we generate synthetic 3D bead images with random aberrations via a realistic image formation model that matches the microscope setup, and we use a simple CNN architecture (which we call \phasenet) to directly predict these aberrations from the given volumetric images.
We demonstrate our approach on two distinct microscopy modalities: \emph{i)} a point-scanning microscope where single-mode aberrations were introduced in the illumination path, and \emph{ii)} a widefield microscope where random-mode aberrations were introduced in the detection path.
We compare the speed and accuracy of \phasenet with the two popular state-of-the-art methods  GS and \Zola, and find that \phasenet leads to better or comparable results yet is orders of magnitudes faster.
Finally, we demonstrate that the number of focal planes required for accurate prediction with \phasenet is related to different symmetry groups of the Zernike modes.

\figOverview

\section{Methods}

Let $h(x,y,z)$ be the acquired image of a bead (point spread function, PSF) and let $\phi(k_x, k_y)$ be the \emph{wavefront aberration}, \ie the phase deviation from an ideal wavefront defined on the back pupil with coordinates $k_x, k_y$.
The wavefront aberration $\phi$ is then decomposed as a sum of Zernike polynomials/modes
\begin{align}
  \phi(k_x, k_y)=\sum\limits_{i} a_i Z_i(k_x, k_y)
  \label{second}
\end{align}
with $Z_i(k_x, k_y)$ being the $i$-th (Noll indexed) Zernike mode and $a_i$ the corresponding amplitude~\cite{bornwolf,Noll1976}.
The problem of phase retrieval is then to infer these amplitudes $a_i$ from $h(x,y,z)$.
Our approach (\phasenet) uses a CNN model that takes a 3D image as an input and directly outputs the amplitudes $a_i$.
Importantly, the model is trained on synthetically created data first and only then applied to real microscopy images~(\cf~\cref{fig:overview}).
That way, we avoid the acquisition of experimental training images with precisely known aberrations, which often is difficult or outright impossible (\eg for sensorless setups).

\subsection{Synthetic training data}
\label{seq:method:synth}

To generate training data for a specific microscope setup, we synthetically create pairs $(a_i^n,h_{synth}^n)_{n \in \N}$ of randomly sampled amplitudes $a_i^n$ and corresponding 3D PSFs $h_{synth}^n$.
We use only the first 11 non-trivial Zernike modes $a^n_i = (a_5^n, \ldots, a_{15}^n)$, excluding \emph{piston}, \emph{tip}, \emph{tilt} and \emph{defocus}, and generate randomly aberrated PSFs by uniformly sampling $a_i^n \in [-0.075,0.075]$.
Given a wavefront $\phi^n = \sum_i a_i^n Z_i$, we compute the corresponding intensity image as:
\begin{equation}
\small
  h_{synth}^n(x,y,z) =| \mathcal{F} \big[ P(k_x, k_y) \cdot e^{2\pi i\phi^n(k_x, k_y)/\lambda} \cdot e^{-2\pi i z\sqrt{n_0^2/\lambda^2 - k_x^2 - k_y^2 }} \big] | ^2
  \label{first}
\end{equation}
where $\mathcal{F}[\cdot]$ is the Fourier transform, $\lambda$ is the wavelength, $n_0$ is the refractive index of the immersion medium, $\phi^n = \sum_{i=5}^{15} a^n_i Z_i(k_x, k_y)$ is the wavefront aberration, and $P(k_x, k_y)$ is the amplitude of the pupil function~\cite{goodman}.
Since we do not consider amplitude attenuation, we simply set $P(k_x, k_y) = \mathbb{1}_{k_x^2+k_y^2 < (NA/\lambda)^2}$ with $\NA$ being the numerical aperture of the objective.
To accommodate for a finite bead size, we then convolve $h_{synth}^n$ with a sphere of appropriate diameter (depending on the experiment) and add realistic Gaussian and Poisson noise.

\figResultsSingle

\subsection{\phasenet}
\label{seq:method:phasenet}

The CNN architecture (\phasenet) is shown in~\cref{fig:overview} and consists of five stacked blocks, each comprising two $\sizethree{3}{3}{3}$ convolutional layers and one max-pooling layer, followed by two dense layers with the last having the same number of neurons as the number of Zernike amplitudes to be predicted ($11$ in our case).
This results in a rather compact CNN model with a total of $0.9$~million parameters.
We use $tanh$ as activation function for all layers except the last, where we use linear activation.
We simulate 3D PSFs $h_{synth}^n$ and the corresponding amplitudes $a_i^n$ which form the input and output of the network, respectively~(\cf~\cref{fig:overview}).
To prevent overfitting, we use a data generator to continuously create random batches of training data pairs during the training process.
We minimize the mean squared error (MSE) between predicted and ground truth (GT) amplitudes and train each model for $50000$ steps and batch size $2$ on a GPU (\textsc{Nvidia} Titan Xp) using the Adam optimizer~\cite{Kingma2014} with learning rate $1 \cdot 10^{-4}$.
Our synthetic training generation pipeline as well as the \phasenet implementation based on Keras~\cite{chollet2015} can be found at \url{https://github.com/mpicbg-csbd/phasenet}.

\figResultsRandom
\subsection{Experimental data}
\label{seq:method:experimental}

We use two different microscope setups (\sted and \wide) to demonstrate the applicability of this technique on real microscopy data.
\paragraph{\sted}

This is a point-scanning microscope designed for STED microscopy, equipped with a $1.4~\NA$ oil immersion ($n_0=1.518$) objective and a $\lambda=755\nm$ illumination laser (\cf~\SF{fig:supp:setup}a and described in~\cite{barbotin2019}).
For these experiments, the system was operated without the STED function activated~–~in effect as a point scanning confocal microscope with open pinhole.
Single Zernike mode aberrations for $Z_5$ (oblique astigmatism) to $Z_{15}$ (oblique quadrafoil) within an amplitude range of $\pm0.11\um$ were introduced in the illumination path via a spatial light modulator (SLM).
The backscattering signal of $80\nm$ gold beads was then measured using a photo-multiplier tube and the stage axially and laterally shifted resulting in $n=198$ aberrated 3D bead images of size $\sizethree{32}{32}{32}$ with isotropic voxel size $30\nm$.
We generated synthetic training data using the given microscope parameters and random amplitudes $(a_5, \ldots ,a_{15})$ in the range of $\pm 0.075\um$~(\cf~\cref{seq:method:synth}).
We then trained a \phasenet model as explained in~\cref{seq:method:phasenet}.

\paragraph{\wide}

This is a custom-built epifluorescence microscope with a $1.1~\NA$ water immersion objective and a $\lambda=488\nm$ illumination laser (\cf~\SF{fig:supp:setup}b).
Mixed Zernike mode aberrations comprising $Z_5 - Z_{10}$ (lower order) or $Z_5 - Z_{15}$ (higher order) were introduced in the detection path via a deformable mirror (DM).
We used an amplitude range of $\pm0.075\um$ for each mode.
The images of $200\nm$ fluorescent beads were recorded at different focal positions, resulting in $n=100$ aberrated 3D bead images of size $\sizethree{50}{50}{50}$ with a voxel size of $86\nm$ laterally and $100\nm$ axially.
As before, we generated similar synthetic training data using the respective microscope parameters and trained a \phasenet model.

\subsection{Evaluation and comparison with classical methods}

We compare \phasenet against two classical iterative methods, GS (Gerchberg-Saxton, code from~\cite{Kner2010}) and \Zola~\cite{Aristov2018}.
GS is an alternating projection method that directly estimates the wavefront aberration $\phi$.
\Zola fits a realistic PSF model to the given image and returns the present Zernike amplitudes~(\SM{sec:supp:class}).
For both GS and \Zola, we used 30 iterations per image, \Zola additionally leveraging GPU-acceleration (\textsc{Nvidia} Titan Xp).
For every method we quantify the prediction error by first reconstructing the wavefront from the predicted Zernike amplitudes (for \phasenet and \Zola) and then computing the root mean squared error (RMSE, in $\um$, averaged over the back pupil) of the difference between the predicted and the ground truth wavefront.

\section{Results}
\figResultsPlanes

\subsection{\sted}

\begin{table}[b]
  \centering
  \begin{tabular}{lrr}
    \toprule
    Method & single ($n = 1$) & batched ($n = 50$) \\
    \midrule
    GS & 0.120~s & 6.2~s \\
    Zola & 17.1~s  & 838~s\\
    \phasenet & 0.004~s & 0.033~s \\
    \bottomrule
  \end{tabular}
  \caption{Runtime of all methods for aberration estimation from a single ($n = 1$) and multiple ($n=50$) PSFs of size $\sizethree{32}{32}{32}$.} 
  \label{tab:timing}
\end{table}

We first investigated the performance of \phasenet on the data from \sted microscope with experimentally introduced single-mode aberrations~(\cf~\cref{fig:results_single}).
This gives us the opportunity to assess the performance of all methods for each Zernike mode and amplitude in isolation.
Here, the respective \phasenet model trained on synthetic PSFs achieved good wavefront reconstruction with the predicted and ground truth wavefront having a median RMSE of $0.025\um$ (compared to a median RMSE of $0.15\um$ for the input wavefronts), thus validating our approach (\cf~\SF{fig:supp:stedsynth}).
We then applied the model on the experimental images, yielding amplitude predictions $(a_5, \ldots, a_{15})$ for each 3D input. 
In~\cref{fig:results_single}a) we show the results for $Z_5$ (oblique astigmatism).
As can be seen, the predicted amplitude $a_5$ exhibits good agreement with the experimental ground truth, even outside the amplitude range used for training (indicated by the gray arrow).
Importantly, the predicted amplitudes for the non-introduced modes ($a_6, \ldots, a_{15}$)  were substantially smaller, indicating only minor cross-prediction between modes~(\cf~inset in~\cref{fig:results_single}a).
The same can be observed for vertical coma $Z_7$ (\cref{fig:results_single}b) and all other modes $Z_6 - Z_{15}$~(\cf~\SF{fig:supp:sted}~\&~\SF{fig:supp:sted2} for reconstructed wavefronts).

We next quantitatively compared the results of \phasenet with predictions obtained with GS and \Zola.
Here, \phasenet achieves a median RMSE between predicted and ground truth wavefronts of $0.028\um$ across all acquired images ($n=198$), which is comparable to the prediction error on synthetic PSFs.
At the same time GS ($0.039\um$) and \Zola ($0.031\um$) performed slightly worse~(\cf~\cref{fig:results_single}c).
This demonstrates that a \phasenet model trained only on synthetic images can indeed generalize to experimental data and achieve better performance than classical methods.  
Crucially, predictions with \phasenet were obtained orders of magnitude faster than with both GS and \Zola~(\cf~\cref{tab:timing}).
Whereas it took only $4ms$ for \phasenet to process a single image, it required $0.12s$ for GS and $17.1s$ for \Zola.
The speed advantage of \phasenet is even more pronounced when predicting batches of several images simultaneously~(\cf~\cref{tab:timing}).

\subsection{\wide}
We next explored the applicability of our approach to the \wide microscope modality, where mixed-mode aberrations were randomly introduced.
The \phasenet model trained on appropriate synthetic data achieved a median RMSE of $0.022\um$ (compared to RMSE $0.14\um$ of the input wavefronts) indicating again good wavefront reconstruction~(\SF{fig:supp:widesynth}).
We then applied the trained model on the experimental bead images.
In~\cref{fig:results_random}a we show results for \phasenet, GS, and \Zola for images with introduced modes $Z_5 - Z_{10}$ (lower order).
The reconstructed wavefronts for both \phasenet and \Zola exhibits qualitatively good agreement with the ground truth, whereas GS noticeably underperforms~(\cf~\SF{fig:supp:widelow}).
Similarly, the calculated RMSE across all images ($n=150$) for GS ($0.124\um$) is substantially larger than for \phasenet ($0.025\um$) and \Zola ($0.012\um$).

The same results can be observed when predicting images with higher order modes $Z_5 - Z_{15}$~(\cref{fig:results_random}b).
As expected, RMSE values increased slightly compared to the lower order modes for all methods, with $0.148\um$ for GS, $0.035\um$ for \phasenet, and $0.019\um$ for \Zola (more examples can be found in \SF{fig:supp:widehigh}).
Although \Zola yields slightly better RMSE than \phasenet for this dataset, \phasenet again vastly outperforms \Zola and GS in terms of prediction time by being orders of magnitude faster~(\cf~\ST{tab:supp:timing2}).

\subsection{Number of input planes}

In both experiments so far, the 3D input of \phasenet consisted of many defocus planes ($n_z=32$ for \sted and $n_z=50$ for \wide).
We set out to determine, whether accurate aberration prediction is still possible with substantially fewer planes.  
We therefore trained several \phasenet models with varying $n_z$ and applied them to experimental images~(\cf~\SM{sec:supp:plane}).
In~\cref{fig:results_planes}a/b  we show predictions with $n_z \in \{1, 2, 32\}$ for single-mode aberrations $Z_5$ (oblique astigmatism) and $Z_7$ (vertical coma).
Interestingly, we find that in the case of $Z_5$ at least $n_z\geq 2$ planes are needed for meaningful predictions, whereas in the case of $Z_7$ already a single plane ($n_z=1$) yields satisfactory results.
This can be explained by observing that for purely $Z_5$ aberrations (\ie $a_{i\neq 5} = 0$), flipping the sign of the aberration amplitude $a_5' =  -a_5$ leads to a 3D PSF that is mirrored along the optical axis. 
Predicting the amplitude $a_5$ from a single image plane is therefore inherently ambiguous.
To further examine this, we grouped the Zernike modes into the classes \emph{even} and \emph{odd} depending on the symmetry of the wavefront (even: $Z_5,Z_6,Z_{11},\ldots$, odd: $Z_7,Z_8,Z_{9},\ldots$) and calculated the prediction for each class separately.
As expected, the RMSE decreases with increasing $n_z$ (\cref{fig:results_planes}c) for both classes.
However, for even Zernike modes the prediction error is significantly higher than for odd modes, especially when using only few planes, in line with our earlier observation.

\section{Conclusion}

We demonstrated that deep learning based phase retrieval with \phasenet using  synthetically generated training data alone does generalize to experimental data and allows for accurate and efficient aberration estimation from experimental 3D bead images.
On datasets from two different microscopy modalities we showed that \phasenet yields better or comparable results than classical methods, while being orders of magnitude faster.
This opens up the interesting possibility of using \phasenet to perform aberration estimation from multiple beads or guide stars across an entire volumetric image in a real-time setting on the microscope during acquisition.
We further investigated how prediction quality depends on the number of defocus planes $n_z$ and found that odd Zernike modes are substantially easier to predict than even modes for the same $n_z$.

Still, our approach may not be applicable to cases where the synthetic PSF model is inadequate for the microscope setup or where experimental data is vastly different from the data seen during training (a limitation that applies to most machine learning based methods). 
Moreover, for discontinuous wavefronts (such as double helix PSFs~\cite{pavani2009} or helical phase ramps~\cite{willig2006}) the low-order Zernike mode representation is likely to be inadequate and \phasenet performance is therefore sub-optimal.
Furthermore, our experimental data so far included only Zernike modes $Z_n \leq 15$, leaving the question open whether our approach would behave similarly for larger Zernike modes.
Additionally, more advanced network architectures that explicitly leverage the physical PSF model might improve prediction accuracy.

We believe that in the future our method can serve as an integral computational component of practical adaptive optics systems for microscopy of large biological samples.

\section*{Funding}
This research was supported by the German Federal Ministry of Research and Education (BMBF SYSBIO II - 031L0044) and by CA15124 (NEUBIAS). MJB and QH were supported by the European Research Council (AdOMiS, no. 695140).  AB was supported by EPSRC/MRC (EP/L016052/1).

\section*{Acknowledgments}
We thank Robert Haase, Coleman Broaddus, Alexandr Dibrov (MPI-CBG) and Jacopo Antonello (University of Oxford) for scientific discussions at different stages of this work.
We thank Nicola Maghelli (MPI-CBG) for valuable input.
We thank Si\^{a}n Culley (UCL, London), Fabio Cunial (MPI-CBG) and Martin Hailstone (University of Oxford) for providing feedback.

\noindent\textbf{Disclosures.} The authors declare no conflicts of interest.

\section*{Supplemental Documents}

See \href{link}{Supplement} for supporting notes, tables, and figures.

\bibliographyfullrefs{sample}

\end{document}